\documentclass{article}
\usepackage[left=1.7cm,right=1.7cm,top=1.5cm,bottom=1.5cm]{geometry}
\usepackage{slashed,color,amsmath,amssymb,graphicx,hyperref,authblk,blkarray}
\usepackage[style=phys,biblabel=brackets,eprint=true,
    pageranges=false,backend=bibtex,sorting=none]{biblatex}
\bibliography{PLB_32879.bib}

\definecolor{myblue}{RGB}{37,52,148}
\hypersetup{
colorlinks = true,   
urlcolor   = myblue, 
linkcolor  = myblue, 
citecolor  = myblue  
}

\title{Left-right model with TeV fermionic dark matter and unification}
\author{Triparno Bandyopadhyay\thanks{Electronic
address:
\href{mailto:gondogolegogol@gmail.com}{\texttt{gondogolegogol@gmail.com}}} ~and
Amitava Raychaudhuri\thanks{Electronic address:
\href{mailto:palitprof@gmail.com}{\texttt{palitprof@gmail.com} }}}
\affil{\small{Department of Physics, University of Calcutta, 92
Acharya Prafulla Chandra Road, Kolkata  700009, India.}}
\date{\today}
\begin{document}
\maketitle
\begin{quote}
    The ingredients for a model with a TeV right-handed scale, gauge
    coupling unification, and suitable dark matter candidates lie at the
    heart of left-right symmetry with broken D-parity. After detailing
    the contents of such a model, with $SU(2)_R$ self-conjugate
    fermions at the right-handed scale aiding in unification of
    couplings, we explore its dark matter implications and collider
    signatures.  
\end{quote}

{\em Introduction:} 
These are indeed exciting times for particle physics as the Large Hadron
Collider (LHC) at CERN is all set to run at its machine configuration of
$\sqrt{s} = 14$ TeV.  With experiments at this highest energy facility
in a hunt for new physics at TeV scales, it is no surprise that the
community is particularly focussed on models with phenomenological
signatures in the $\mathcal{O}$(TeV) range. Of the various models
that try to explain natural phenomena beyond the scope of the standard
model (SM), those based on left-right (LR)  symmetry \cite{lrs1, lrs2,
lrs3, lrs4} have withstood the tests of time as they extend the SM
electroweak sector in well motivated ways.  These models explain the
origin of parity violation and at the same time gauge the global
$U(1)_{(B-L)}$ symmetry inherent in SM and in the process explain the
smallness of the neutrino mass.

Hypothesised primarily in the context of visible sector physics, LR
models do not have any {\em de facto} dark matter (DM) candidate built
into their bare bones structure.  However, the group theoretic
configuration of LR symmetry has the provision of a naturally arising
discrete symmetry, remnant after the spontaneous breaking of
$U(1)_{(B-L)}$ \cite{disc1, disc2, disc3, disc4, disc5,disc7}, which
facilitates the building of a plethora of DM models
\cite{netdm1,netdm2,dmlrs3,dmlrs4,dmlrs11,dmlrs7}.

The LR gauge symmetry and particle content, along with gauge coupling
unification (GCU), can be embedded in $SO(10)$ ``grand unified
theories'' (GUTs) \cite{so101,so102} having numerous desired features
such as quark-lepton unification, unification of the SM interactions,
and explanation of the arbitrary $U(1)_Y$ assignment of the SM, among
others. However, in models with the left-right symmetry breaking scale
$M_R$ $\sim\mathcal{O}$(TeV), and a minimal scalar sector, GCU is
impossible \cite{5auth,runlrs2,runlrs1,runlrs3,jdarc1,runlrs4}. To achieve
unification one either needs to add scalar multiplets redundant to their
primary function of symmetry breaking and mass generation, or larger
symmetries intermediate between the Left-Right symmetry (LRS) and GUT
scales.  These modifications end up introducing additional scalar fine
tunings and a degree of arbitrariness.

In this letter, we show that the three requirements of
$\mathcal{O}$(TeV) right-handed breaking scale, unification of LRS
couplings, and the presence of a  suitable dark matter candidate can be
achieved with a single stroke by the careful appraisal of fermion masses
in a class of left-right models where the exact $L\leftrightarrow R$
symmetry is spontaneously broken at a scale different from the one where
the right-handed gauge symmetry is broken \cite{dp1, dp2}.  While
focussing on model mechanics, we discuss dark matter phenomenology and
show that though its direct detection prospects are not bright, the
collider signatures of the model are testable.

{\em Model:}
The left-right symmetry  is defined by the gauge group, $SU(3)_C \times
SU(2)_L\times SU(2)_R\times U(1)_{(B-L)}$, and a discrete
$SU(2)_L\leftrightarrow SU(2)_R$ symmetry, $\mathcal{P}$. Under this,
the SM quarks, leptons, and a right-handed (RH) neutrino of one family
transform as:
\begin{align}
    l_L &\equiv (1_C,2_L,1_R,-1_{(B-L)});\, {l}_R \equiv
    (1_C,1_L,2_R,-1_{(B-L)});\nonumber \\
    q_L &\equiv (3_C,2_L,1_R,1/3_{(B-L)});\, {q}_R \equiv
    (3_C,1_L,2_R,1/3_{(B-L)})\,;
\end{align}
with $(B-L)$ being normalised by the  relation:
\begin{align}
    Q_{em}&=T_{3R}+T_{3L}+\frac{B-L}{2} \;\;.
\end{align}
The scalar sector is given by: 
\begin{align}
    \Phi &\equiv
    (1_C,2_L,2_R,0_{(B-L)});\,\eta\equiv(1_C,1_L,1_R,0_{(B-L)});\nonumber \\
    \Delta_R &\equiv
    (1_C,1_L,3_R,2_{(B-L)});\, \Delta_L\equiv (1_C,3_L,1_R,2_{(B-L)}) 
    \;\;. 
\end{align}
Under $\mathcal{P}$ the multiplets transform as:
\begin{align}
    l_L\leftrightarrow l_R;&\qquad q_L\leftrightarrow q_R; \qquad
    \Delta_L\leftrightarrow
    \Delta_R;\qquad \Phi\leftrightarrow \Phi^\dagger;\qquad \eta
    \leftrightarrow -\eta \;\;. \label{eq:trans} 
\end{align}
The scalar sector is modified to accommodate spontaneous breaking of
$\mathcal{P}$ at a scale $M_\mathcal{P}$, where the $\mathcal{P}$ odd
gauge singlet, $\eta$, acquires a vacuum expectation value ($vev$),
$v_\eta$. Thus, symmetry breaking takes place in three steps. The first
being the breaking of $\mathcal{P}$, followed by the breaking of
$SU(2)_R \otimes U(1)_{(B-L)}$  to SM by the $vev$, $v_R$, of
$\Delta_R$, and finally electroweak symmetry breaking is achieved
through the $vevs$ $k_1$ and $k_2$ of the bi-doublet $\Phi$, with
$\sqrt{k_1^2+k_2^2}=246$ GeV. We show that the other mass scales of the
model are $M_R \sim \mathcal{O}$(TeV) and $M_\mathcal{P}$ at the GUT
scale.

The gauge bosons related to $SU(2)_R\otimes U(1)_{(B-L)}$ breaking,
$W_R^\pm$ and $Z'$, acquire mass at $M_R$, and for
$\left({M_W}/{M_{W_R}}\right)^2\ll 1$ are given by:
\begin{align}
    M_{W_R} &= \frac{g_R}{\sqrt{2}}v_R \;\;;\qquad M_{Z'} =
    \frac{\sqrt{2}}{{\cos\phi}} ~M_{W_R} \;\;,\\
    \text{with}\qquad \sin\phi &= \frac{g_L}{g_R} \tan\theta_W\nonumber
    \;\;,
\end{align}
where $\theta_W$ is the weak mixing angle. For
$\left({M_W}/{M_{W_R}}\right)^2 \ll 1$, $W_L$--$W_R$ mixing is
negligible. The physical states of the bi-doublet other than the SM
Higgs are constrained to be $\geq \mathcal{O}($10 TeV) from lepton
flavour violation limits \cite{bdb1}, and the scalars from $\Delta_L$
are all heavy at $M_\mathcal{P}$ \cite{dp1, dp2}. There are no stringent
constraints on the masses of the $\Delta_R$ scalars and they can be
lighter than $M_{W_R}$  and even $\mathcal{O}($100 GeV). Here, for
simplicity we take them to be heavier than $M_{W_R}$.

$U(1)_{(B-L)}$ being broken by a scalar with $(B-L)$ = 2, leaves
behind a remnant $\mathbb{Z}_2$ symmetry, defined by:
$\mathcal{Z}\equiv (-1)^{3(B-L)}$ \cite{disc5}.  LRS fermions
(scalars) have odd (even)  $3(B-L)$ and hence are odd (even)
under $\mathcal{Z}$. As a result, fermions with even $3(B-L)$ are
forbidden to decay only to SM fermions and/or bosons, and hence
the lightest one of them is stable. If this state is neutral,
then subject to relic density and direct/indirect detection
constraints, it can be taken to be a dark matter candidate.

Fermions in self-conjugate representations of $SU(2)_L\otimes SU(2)_R$,
$X_L \oplus X_R\equiv (1_C, (2m+1)_L, 1_R, 0_{(B-L)}) \oplus
(1_C,1_L,(2m+1)_R,0_{(B-L)})$, typify this scenario with $m\in
\mathbb{N}$. Each multiplet consists of a Majorana fermion and $m$ pairs
of Dirac fermions and their antiparticles with electric charges $1$ to $m$.
Thus, these multiplets must be assigned $B=L=0$.

The left-right symmetric bare mass and Yukawa terms of these multiplets
for a general case of $n_g$ such `generations' is given by:
\begin{align}
    \mathcal{L}_{X_M} &= \frac{\mathcal{M}_i}{2}\left( \overline{{X^i_L}^c} X^i_L
    + R\leftrightarrow L\right ) +  \frac{h_i}{2}(v_\eta
    +\eta)\left(\overline{{X^i_L}^c}X^i_L - R\leftrightarrow L \right) 
      +  h.c.\; ,\label{eq:1}
\end{align}
where summation over $i$= $1,\cdots,n_g$ is implicit.
The negative signs pertaining to interactions of $X_R^i$ with $\eta$ are
according to eq. [\ref{eq:trans}]. Because the multiplets can
always be rotated into a diagonal basis, we can do away with the cross
terms without any loss of generality. From eq. [\ref{eq:1}], we see that
the breaking of $\mathcal{P}$ enforces a separation of the
masses of the  multiplets transforming under $SU(2)_L$ and
$SU(2)_R$ with the corresponding masses given by:
\begin{align}
    M_i^L &= \mathcal{M}_i + h_i v_\eta;\;
    M_i^R (=:M_i) = \mathcal{M}_i - h_i v_\eta\, .
\end{align}
With $\mathcal{M}_i \sim h_i v_\eta$, $X_L^i$ multiplets remain heavy at
$M_\mathcal{P}$ and the $X_R^i$ become light with the exact mass scale
dependent on the couplings. We want to underscore that in general for
$(1_C, (2m+1)_L, 1_R, 0_{(B-L)}) \oplus (1_C,1_L,(2m+1)_R,0_{(B-L)})$
fermion multiplets, the mass scale of either one will be at the larger of
$\mathcal{M}_i$ and $h_i v_\eta$ while the other can be tuned to be at
lower values.  During the evolution of the Universe the superheavy
$SU(2)_L$ multiplets are Boltzmann suppressed and annihilate and
co-annihilate rapidly to lighter states through their couplings to $W_L$
and $Z$.

The framework being discussed can lead to a variety of DM models, which
we label as $(m,n_g)$, with the DM particle(s) completely separated from
the SM and interacting only with the RH sector.  For the rest of the
letter we focus on the (1,2) case as this model simultaneously provides
a suitable DM sector, gauge coupling unification, and 
$\mathcal{O}$(TeV) $M_R$. 

{\em Gauge Coupling Unification:}
With the self-conjugate $SU(2)_R$ generations of an $(m, n_g) \equiv (1,2)$ model,
i.e., the model with a pair of $(1_C,3_L,1_R,0_{(B-L)})+
(1_C,1_L,3_R,0_{(B-L)})$, in the TeV range, the gauge couplings unify
with $M_\mathcal{P}=M_U$. 
The LR gauge group is a subgroup of $SO(10)$ and with GCU we can
embed the model in an $SO(10)$ unified theory. The LRS multiplets
of the model belong in the following $SO(10)$ representations:
\begin{align}
&(3_C,2_L,1_R,1/3_{(B-L)})+(\bar{3}_C,1_L,2_R,-1/3_{(B-L)})
+(1_C,2_L,1_R,-1_{(B-L)})+(1_C,1_L,2_R,1_{(B-L)})\subseteq
16_F;\nonumber \\ & (1_C,2_L,2_R,0_{(B-L)})\subset 10_H; \;
(1_C,3_L,1_R,2_{(B-L)})+(1_C,1_L,3_R,2_{(B-L)})\subset 126_H;\;
(1_C,1_L,1_R,0_{(B-L)})\subset 210_H; \nonumber\\
&(1_C,3_L,1_R,0_{(B-L)})+(1_C,1_L,3_R,0_{(B-L)}) \subset 45_F;
\end{align}
where the subscripts $F$ and $H$ denote whether the
multiplets contain fermions or scalars, respectively. There is an
element of the $SO(10)$ algebra, `$\mathcal{D}$' \cite{sln},
which in the case that all the couplings of the lagrangian are
real, plays the role of the parity symmetry $\mathcal{P}$.
$\eta\subset 210_H$ is odd under `$\mathcal{D}$'.

The fermion  triplets reside in 45-plets. The $SO(10)$ symmetric mass
term for which is: 
\begin{align} 
    \mathcal{L}_{\text{Mass}} &= -\frac{\mathcal{M}_{1,2}}{2} 
    \overline{{45_F^{1,2}}^C}45_F^{1,2} + h.c.\;\;, 
\end{align} 
with $\mathcal{M}_{1,2}  \sim M_U = M_\mathcal{P}$.
Under the Pati-Salam \cite{lrs1,lrs2} symmetry, $SU(4)_C\otimes
SU(2)_L\otimes SU(2)_R$, $45_F$ is decomposed as: $45 \supset
(15_4,1_L,1_R) + (6_4,2_L,2_R) + (1_4,3_L,1_R) + (1_4,1_L,3_R)$.  Since
$(15_4,1_L,1_R)$ and $(6_4,2_L,2_R)$ transform identically under
$SU(2)_L$ and $SU(2)_R$, they have masses at $\mathcal{M}_{1,2}$, while
$(1_4,3_L,1_R)$ and $(1_4,1_L,3_R)$ are split according to the previous
discussion. As for the scalars, all submultiplets not 
required to be either at the right-handed or the electroweak scale
are at the unification scale according to the minimal fine-tuning principle
of the extended-survival hypothesis \cite{esh1,esh2}.

\begin{figure}[thb]
    \centering
    \includegraphics[width=0.5\textwidth]{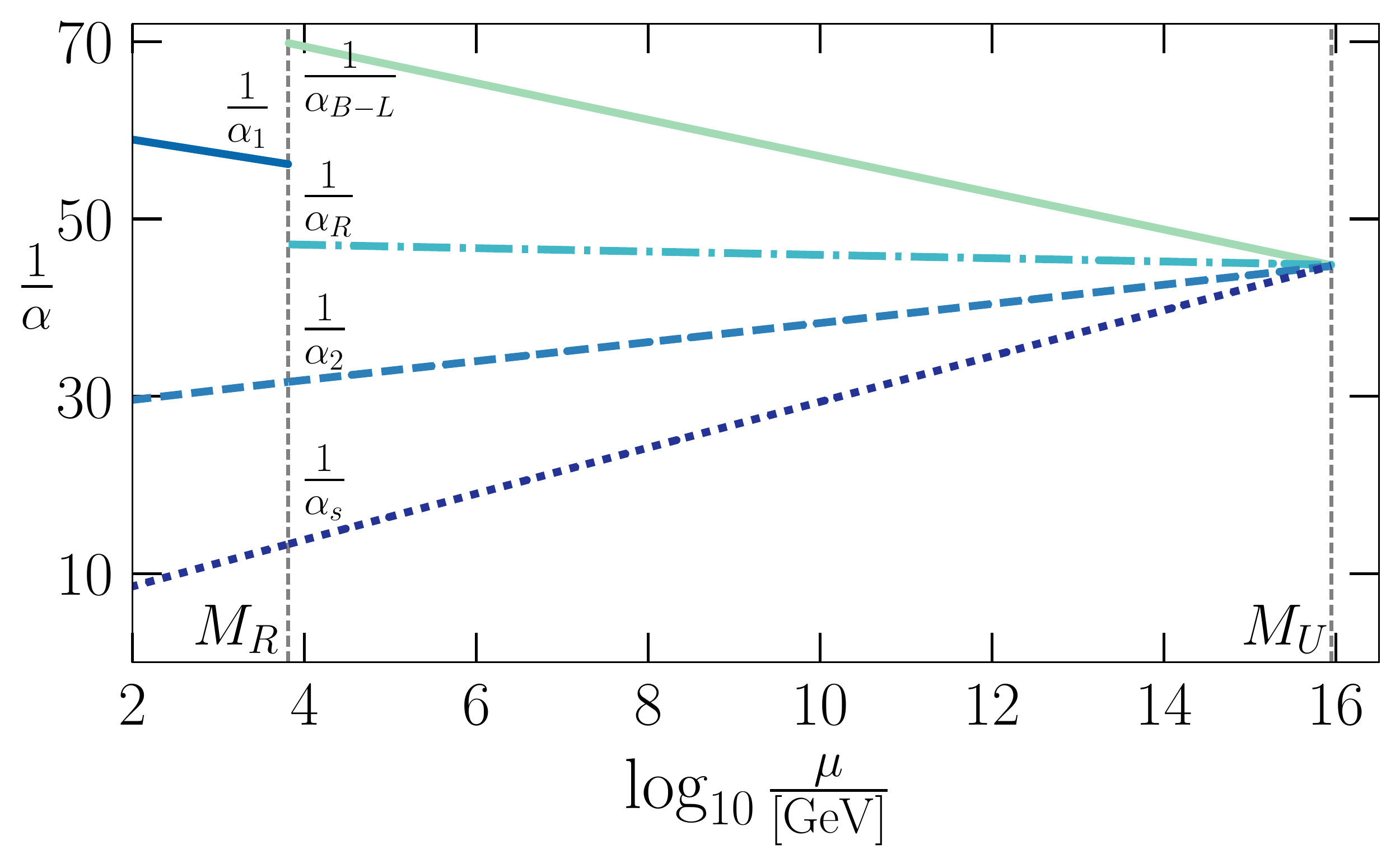}
    \caption{2-loop running of the inverse of the gauge couplings
    $\alpha_i$ with mass scale, $\mu$.}
    \label{fig:running}
\end{figure}

In Fig. [\ref{fig:running}], we show the running of the inverses of the
fine structure constants ($\alpha=g^2/(4\pi))$,
as obtained from 2-loop perturbation theory. As inputs at the Z-pole,
$M_Z=91.1876(21)$, we take $\alpha_s=0.1181(11)$,
$\sin^2\theta_W=0.23129(5)$, and $\alpha_{EM} = 1/128$ \cite{rpp}. We find
that when $\alpha_s(M_Z)$ and $\sin^2\theta_W$ are varied over their
1$\sigma$ allowed ranges the unifcation scale varies between $(0.81
-1.05) \times 10^{16}$ GeV and the unification coupling comes out to be,
$g_U=0.53$. The $SU(2)_R$ breaking scale lies between 3.78 -- 9.40 TeV,
with the $SU(2)_R$ coupling, $g_R$ = 0.52. Fig.  [\ref{fig:running}] has
been drawn using the central value. The $U(1)$ couplings of the theory
are normalised according to GUT (canonical) normalisation, resulting in
the matching condition at $M_R$:
\begin{equation}
    \frac{1}{g_Y^2}=\frac{3}{5}\frac{1}{g_R^2} +
    \frac{2}{5}\frac{1}{g_{\text{(B-L)}}^2} + \frac{1}{20\pi}
\end{equation}
In between $M_R$ and $M_U$ the particles flowing in the loops and hence
contributing to the $\beta$-coefficients are $\Phi$, $\Delta_R$, $l_L$,
$l_R$, $q_L$, $q_R$ as in traditional D-parity broken models, and the
pair of dark sector $SU(2)_R$ triplets $X^{1,2}_R$. The system of
running equations are given by \cite{mcv,jones}:
\begin{align}
    \frac{\partial g_i}{\partial\log\mu}&=
    \frac{a_i}{16\pi^2}g_i^3+\sum_j\frac{b_{ij}}{(16\pi^2)^2}g_i^3g_j^2
\end{align}
The 1-loop $\beta$-coefficients $a_i$ and the 2-loop
$\beta$-coefficients, $b_{ij}$, for the couplings of the SM are readily
available \cite{jones}, the same for the LRS stage are given in Eq.
[\ref{eq:beta}].  Since the only additions on top of the usual LRS
particle content are the self-conjugate $SU(2)_R$ triplets which
transform trivially under the other symmetries, the only change in the
$\beta$-coefficients are for the $SU(2)_R$ coupling for the 1-loop case
and the diagonal coefficient corresponding to $SU(2)_R$ for the two loop
case.
\begin{align}
    &\begin{blockarray}{rccccl} 
        &B-L&2R&2L&3C&\\
            \begin{block}{r[cccc]l}
                a_i\equiv &\frac{11}{2}&\frac{1}{3}&-3&-7&\\
            \end{block}\\
            \begin{block}{r[cccc]l}
                b_{ij}\equiv & \frac{61}{2}&\frac{81}{2}&\frac{9}{2}&4&B-L\\
                \\
               & \frac{27}{2}&\frac{208}{3}&3&12&2R\\
                \\
               & \frac{3}{2}&{3}&8&12&2L\\
                \\
                & \frac{1}{2}&\frac{9}{2}&\frac{9}{2}&-26&3C\\
            \end{block}
    \end{blockarray} \label{eq:beta}
\end{align}

In principle, a complete treatment of 2-loop RGE running should take
into account threshold effects \cite{trs1,trs2} at all the symmetry
breaking scales. However, in this work we do not include threshold
effects, as in demanding exact unification of the couplings, we are
establishing GCU for a more restricted case. Threshold
corrections introduce more parameters to the model and hence such
situations are bound to follow suit.  

We next estimate the lifetime of the proton in our model. 
In non-supersymmetric GUTs, scalar induced $d=6$ and the $d>6$ operators
contributing to proton decay are generally highly suppressed in
comparison to the gauge induced $d=6$ operators \cite{pd2,pd3,pd4}, and
here we concentrate only on the latter. The decay rate of the proton in
the $p\rightarrow e^+\pi^0$ channel is expressed as \cite{pd1,pd2}:
\begin{equation}
    \Gamma(p\rightarrow e^+\pi^0) = \frac{ m_p g_U^4}{
    16\pi f^2_\pi M_U^4}
    R_L^2(A_{SL}^2+A_{SR}^2)|\alpha_H|^2(1+D+F)^2 
\label{pdecay}
\end{equation} 
where $m_p$=938.3 MeV \cite{rpp} is the mass of the proton,
$f_\pi$=130.41(23) MeV \cite{pdc} is the pion decay constant,
$\alpha_H=-0.0118(0.0021)$ GeV$^3$ denotes the relevant hadronic matrix
element, $D=0.8(2)$ and $F=0.47(1)$ are chiral lagrangian parameters
calculated from lattice gauge theory \cite{as1,as2,as3}. $g_U$ is the
unified coupling constant, $M_U$ the unification scale.  $R_L$=1.46 is
the two-loop long range running effect on the effective proton decay
operator, corresponding to running from $M_Z$ to $m_p$, while
$A_{SL(R)}$ is the short range left-(right-) handed short range
renormalisation factor of the proton decay operator corresponding to
running from $M_U$ to $M_Z$ \cite{pdrgl}.  $A_{SL(R)}$ is a function of
the anomalous dimensions and $\beta$-coefficients of the running
couplings, and also the values of the couplings at the symmetry breaking
scales and are taken to be $A_{SL} \simeq A_{SR} = 2.0$
\cite{pdals1,pdals2,pdals3,pdals4,pdals5}. We set the masses of the
leptoquark gauge bosons to be degenerate and at $M_U$.  Further, the
flavor matrices associated with baryon and lepton flavor changing
currents have been set to unity \cite{pd5,pd6}.  

With $M_U = 10^{15.97}$ GeV we get from eq.  [\ref{pdecay}]  a proton
decay lifetime in this channel, $\tau_{p\rightarrow e^+\pi^0} \sim
1.5\times 10^{35}$ years, which is larger than the present bound of
$\tau_{p\rightarrow e^+\pi^0}=1.6\times 10^{34}$ years \cite{skk}, but
testable at the Hyper-Kamiokande experiment \cite{hkk}, which is
expected to probe lifetimes $\sim 2\times 10^{35}$ yrs. As indicated by
eq. [\ref{pdecay}], this value is extremely sensitive to the unification
scale $M_U$. Still, we have checked that for the above chosen values of
the parameters, $\tau_{p\rightarrow e^+\pi^0}$ remains below the
Hyper-Kamiokande projection with $M_U$ varying between its allowed
range, i.e., (0.81 -- 1.05) $\times 10^{16}$ GeV. However, extreme
choices of the different parameters may make the model not falsifiable
even by this experiment.

The mass scales, predicted by unification, are
particularly ingratiating for neutrino seesaw masses. In minimal LR
models, the left-handed neutrino has both type-I and type-II seesaw
\cite{ss1,ss2, ss3,ss4} contributions. L $\leftrightarrow R$ symmetry
breaking induces a nonzero $SU(2)_L$ triplet $vev$ \cite{sosesa3}:
\begin{equation} 
    v_L\simeq  \frac{v_R}{v_\eta} ~\frac{{\mathcal
O}(k_i^2)}{2M_{\eta\Delta}} \;\;.
\end{equation} 
Here $M_{\eta\Delta}$ is the dimensionful coefficient of the
$\eta\Delta\Delta$ type term in the potential.  The left- and
right-handed neutrino masses are given by \cite{sosesa1, sofermas1,
sosesa2}: 
\begin{align} 
    M_{\nu_R} &= f v_R \quad , \quad
    M_{\nu_L} = f v_L + \frac{v^2}{v_R} yf^{-1}y^T \;\;.  
\end{align} 
$f$ is the Yukawa coupling matrix of the leptons with the triplet
scalars $\Delta_{L,R}$ while $y$ is the Yukawa matrix of the
neutrinos with the bidoublet $\Phi$.  From the values of the
symmetry breaking scales as given above, we see that the
left-handed neutrino gets a mass of the order of 0.1 eV, with $f
\sim \mathcal{O}(1)$, when the Yukawa matrix $y$ is set at the
order of that of the up quark, in the spirit of quark-lepton
unification\footnote{Grand unification implies the same Yukawa couplings
for up-type quarks and the neutrinos. However, the contributions to the
masses in the two sectors can be the same, for the $(1_4,2_L,2_R)
\subset 10_H$, or unequal and of opposite sign, for $(15_4,2_L,2_R)
\subset 126_H$. For the second and the third generation a fine-tuned
cancellation between the two contributions (at the level of 1 in
$10^{5}$ for the third generation)  is needed to keep the Type I seesaw
neutrino masses in the desired range.}. The seesaw is predominantly type
I.

{\em Dark Matter Phenomenology:}
The triplets, $X_R^{1,2}$, each contribute a singly-charged Dirac
fermion--anti-fermion pair ($\chi^\pm_{1,2}$), and a Majorana fermion
($\chi^0_{1,2}$). The charged and neutral states are mass degenerate at
tree level, with mass $M_{1,2}$. At one-loop order, gauge interactions induce 
the mass splitting, $\Delta_M^{1,2}=M^{1,2}_{\chi^\pm}-M^{1,2}_{\chi^0}$
\cite{msp1,msp2,mdm1}.  

The interaction lagrangian for the constituents of the
triplets, $X_R^{1,2}$, for the LR stage is given by:
\begin{align}
    \mathcal{L}_{\text{int}} &= - g_R
    \left(\overline{\chi^+_i}\slashed{W}_{R}^+\chi^0_i +h.c.\right)
    - e \overline{\chi^+_i}\slashed{A}\chi^+_i - g_R\cos\phi_0
    ~\overline{\chi^+_i}\slashed{Z}'\chi^+_i + e\tan\theta_W
    \overline{\chi^+_i}\slashed{Z} \chi^+_i \label{eq:lag_int}
\;\; (i = 1,2) \;\;,
\end{align}
where $e$ is the electromagnetic coupling.
\begin{figure}
    \centering
    \includegraphics[width=0.5\textwidth]{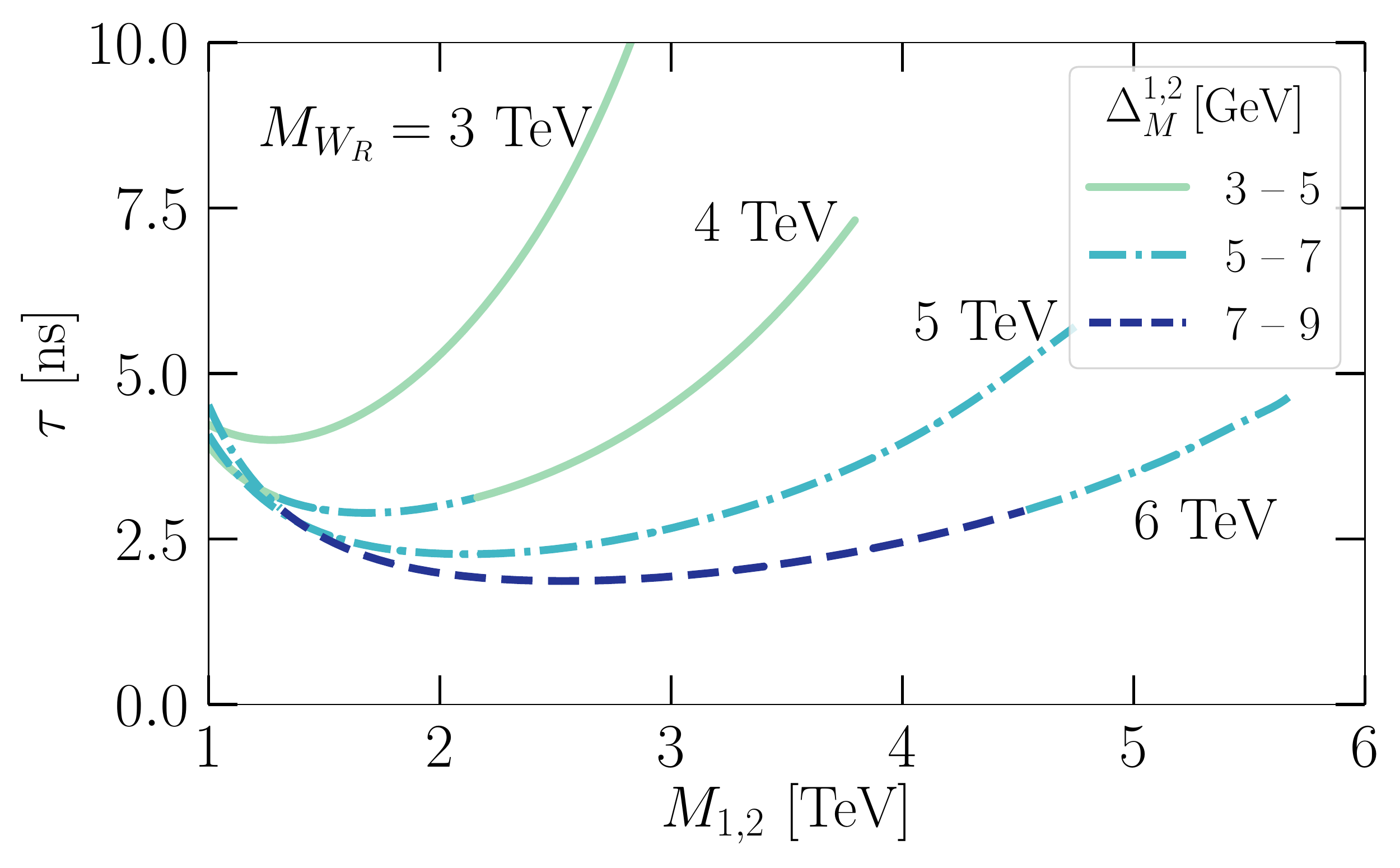}
    \caption{The decay time of $\chi_{1,2}^\pm$
as a function of its mass for different choices of $M_{W_R}$. 
Linestyles distinguish different mass splittings between the charged and
neutral states of the multiplet. }
    \label{fig:lifet}
\end{figure}
Presence of charged heavy fermions during big bang nucleosynthesis (BBN)
would imply the existence  of atom-like bound states, in the present
epoch, containing such particles \cite{hvp1,hvp2}.  The non-observance
of such entities in deep sea water searches \cite{hh1,hh2,hh3,hh4,hh5}
rules out their existence. This implies positive $\Delta^{1,2}_M$, large
enough to produce a lifetime for the charged states smaller than the
time at BBN $\sim 1$ s.

In Fig. [\ref{fig:lifet}] we plot the decay time of $\chi_{1,2}^\pm$ as
a function of its mass for different $M_{W_R}$ near $M_R$. The
intra-multiplet mass splitting, $\Delta_M^{1,2}$, calculated using
expressions in \cite{dmlrs3,dmlrs4}, is indicated by the line-styles of
the curves, i.e., short-dashed, long-dashed, or solid.  Notice that for
each curve $\Delta_M^{1,2}$ changes with $M_{1,2}$.  We find that the
lifetime of the charged states for all masses near $M_{W_R}$ is
$\mathcal{O}$(ns), and the mass splitting is $\mathcal{O}$(GeV). Hence
the heavy charged states of our model decay well before BBN. As the mass
difference is  tiny with respect to the masses themselves,
$\chi_2^\pm\rightarrow \chi_2^0\chi_1^\pm\chi_1^0$ decay is forbidden
from kinematics. Of course the same argument also applies the other way
round. Hence, although we have a single stabilising $\mathbb{Z}_2$, we
end up with two component ($\chi_{1,2}^0$) dark matter.

\begin{figure}[th]
    \centering
    \includegraphics[width=0.5\textwidth]{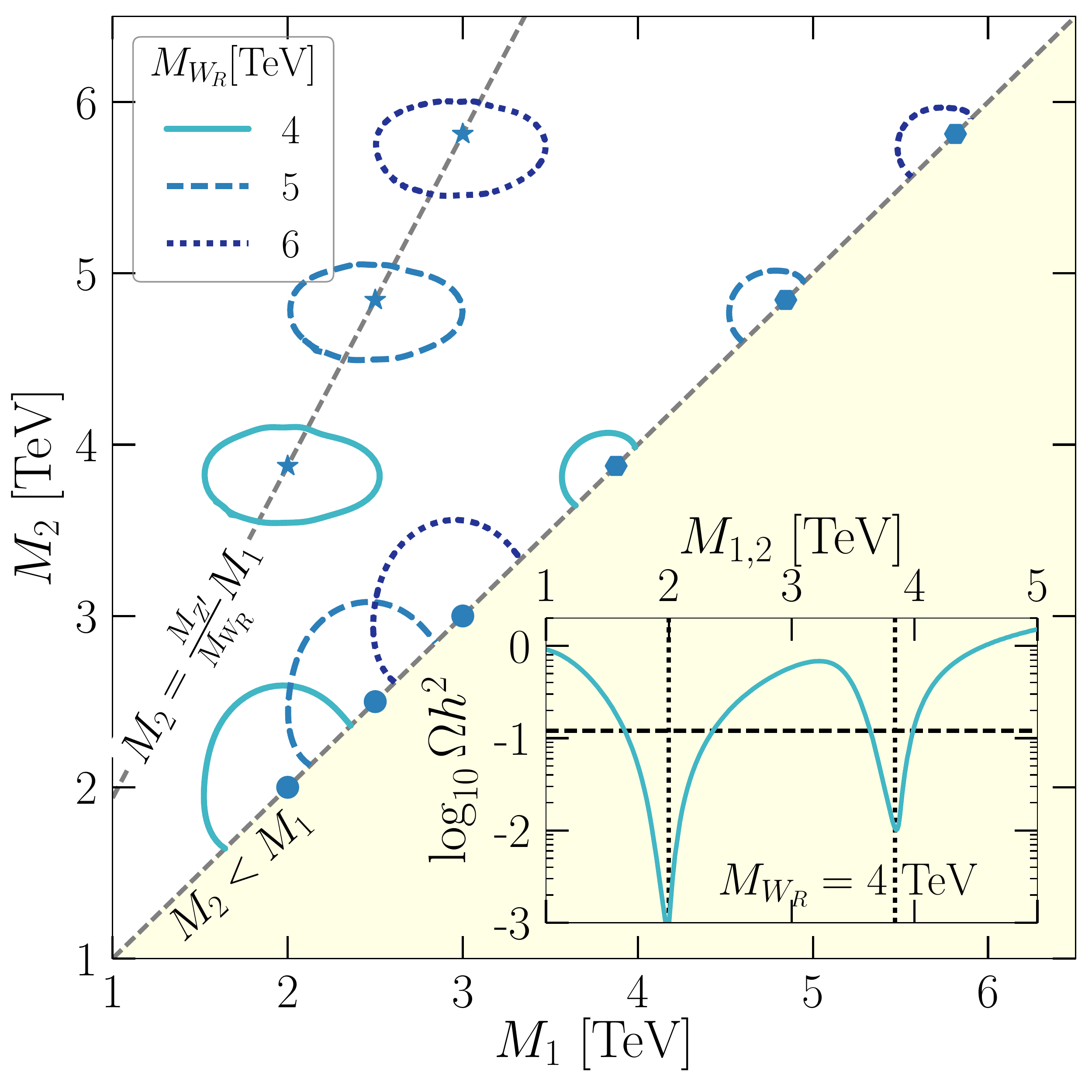}
\caption{The points in the $M_1-M_2$ plane consistent
with the measured dark matter relic density lie on the
curves (see text). Only the solutions with $M_2 > M_1$ are
displayed. Plots are shown for different $M_{W_R}$.
Inset: The dark matter relic density as a function of its mass
for $M_{W_R} = 4$ TeV. The curve is for the case when
the two dark matter candidates are degenerate.  }
    \label{fig:relden}
\end{figure}

The behaviour of dark matter relic density for this model is illustrated
in Fig. [\ref{fig:relden}]. The allowed regions in the $M_1 - M_2$ plane
are those points which fall on the ellipse-like  or semi-circle-like plots. We show only
the region for which $M_1 < M_2$. The allowed values with $M_1 > M_2$
can be readily obtained by a reflection. In the inset of Fig.
[\ref{fig:relden}] we exhibit the relic density as a function of the
dark matter mass for $M_{W_R} =$ 4 TeV. The observed value of the relic
density, $\Omega h^2 = 0.1198\pm 0.0015$ \cite{dmplanck},  is indicated
by the dashed horizontal straight line. As noted, in the model under discussion,
there are two dark matter candidates, $\chi_1^0$ and $\chi_2^0$. In the
inset, for simplicity, they have been taken to be degenerate.  The dips
in the curve reflect resonant $\chi_i^\pm \chi_i^0 \rightarrow W_R^\pm$
or $\chi_i^+ \chi_i^- \rightarrow Z'$ production. Without these dips,
the relic density in this model would have been about an order of
magnitude larger than the observed limit. The points where the curves
agree with the observation are near the two resonant dips. In Fig.
[\ref{fig:relden}] the closed ellipse-type curves with an asterisk
in the middle correspond to regions where the dark matter candidate
$\chi_2^0$ is near the $Z'$ resonance (i.e., $M_{2} \simeq M_{Z'}/2$)
while $\chi_1^0$ is close to the $W_R$ resonance point (i.e., $M_{1}
\simeq M_{W_R}/2$). The semicircle-like curves with a dot
(hexagon) within correspond to the situation where the dark matter
particles $\chi_1^0$ and $\chi_2^0$ are near degenerate and also close
in mass to $M_{W_R}/2$ ($M_{Z'}/2$). We have kept the lower bound of
$M_{1,2}>547$ GeV, as set by recent searches for heavy singly charged
particles \cite{hsp1,hsp2}.  For these relic density computations we
have utilized the MicrOMEGAS 4.3 \cite{micromegas} package. The model
file was written using FeynRules 2.0 \cite{feynrules}, modifying the
version in \cite{feynlrs} to our needs.

At freeze-out temperature, the charged states, $\chi_{1,2}^\pm$, did not
have enough time to decay to the neutral ones, and hence annihilation
and co-annihilation of all the triplet states contribute to the net
annihilation cross section $\langle\sigma v\rangle$. Near the $W_R$
resonance, $\langle\sigma v\rangle$ is saturated by co-annihilation of
$\chi_{1,2}^\pm$ with $\chi_{1,2}^0$ and around the $Z'$ resonance by
both co-annihilation and annihilation of $\chi_{1,2}^\pm$.  As the
neutral $\chi_{1,2}^0$ have no interaction with the $Z'$ or $Z$, it can
only annihilate to a $W_R$ pair through the $t$-channel exchange of
$\chi_{1,2}^\pm$. This channel, however, opens up only when $M_{1,2}
\gtrsim M_{W_R}$, and even then it accounts for a minute fraction of the
total $\langle\sigma v\rangle$.  $g_R$ is essentially fixed from the
running of gauge couplings and is no more a free parameter while
calculating cross sections. Furthermore, the relic density constraint
fixes a narrow range for $M_{1,2}$ given $M_{W_R}$.  The scale of
$M_{W_R}$ itself is fixed by gauge coupling unification. This makes the model
remarkably predictive and free of parameters which can be altered at
will. Thus, falsifying the model is quite straightforward. 

Present and proposed DM direct detection experiments such as LUX,
LZ, XENON1T \cite{dirdlux,dirdlz,dirdx1t}, are all based on
detecting elastic scattering of WIMP DM candidates with nucleons.
The dark matter candidates of this model, $\chi_{1,2}^0$, do not have
any neutral current interactions, neither do they couple to the
Higgs boson. Their only possible interaction with nucleons
($\mathcal{N}$) are through charged current processes, $\chi_i^0
\mathcal{N}^0 \rightarrow \chi_i^- \mathcal{N}^+$ or $\chi_i^0
\mathcal{N}^+ \rightarrow \chi_i^+ \mathcal{N}^0$. At the direct
detection experiments, the $\mathcal{N}$ is initially at rest and
the DM kinetic energy alone is not large enough to surmount the
$\mathcal{O}($GeV) mass difference between the $\chi_{1,2}^\pm$ and
$\chi_{1,2}^0$. Therefore, an on-shell $\chi_{1,2}^\pm$ in the final
state is disallowed from kinematic considerations.  An off-shell
$\chi_{1,2}^\pm$ decaying to $\chi_{1,2}^0$ and $(l ~\nu_l)$ or pions
through $W_R^*$ is in principle possible but highly suppressed
due to lack of available phase space and $\mathcal{O}($TeV)
masses in the propagators. Neutral current NLO cross sections for
$\chi_i^0 \mathcal{N}\rightarrow \chi_i^0 \mathcal{N}$, involving
$W_R$ and $\chi_{1,2}^\pm$ in the loop are naturally negligible. A
detailed discussion in case of $SU(2)_L$ triplets and a
possible way of circumventing the difficulty in detection can be
found in \cite{mdm1}.   In the absence of any
annihilation channels at tree level, the DM parameter region is not
constricted by indirect detection \cite{hess1,hess2} constraints.

{\em Collider Studies:}
As noted previously, the dark matter relic density constraint restricts
the masses of $\chi_{1,2}^\pm$ and $\chi_{1,2}^0$ to near $M_{W_R}/2$ or
$M_{Z'}/2$. The $\chi_{1,2}^\pm$ particles, if produced, for example,
through  $W_R$ or $Z'$ decay,  will be observed as tracks in the CMS and
ATLAS pixel detectors and silicon trackers.  These particles will
typically be at sub-relativistic velocities and can be distinguished
from SM charged particles from the higher rate of ionization energy loss
($dE/dx$).  For most of the allowed mass region, the final state
particles have $0.3<\beta\gamma (=p/M)<1.5$ and hence the average energy
loss with distance travelled can be modelled by the Bethe-Bloch
distribution .  Given a lifetime of $\mathcal{O}($ns) for
$\chi_{1,2}^\pm$, as can be seen from Fig. [\ref{fig:lifet}], we find
their decay lengths to be of the order $\sim 0.1$ -- 1 m.  The charged
particles will hence decay almost exclusively in the trackers of CMS and
ATLAS.  The only decay mode of $\chi_{1,2}^\pm$ is to $\chi_{1,2}^0$,
and the mass difference being $\sim \mathcal{O}($GeV), the associated
jets will be too soft to be reconstructed for a displaced vertex
analysis. Hence, the signal of the charged particles will be the
observation of disappearing tracks\footnote{ For a recent discussion of
the sensitivity of the LHC detectors to such disappearing charged
tracks, see for example, \cite{distrac}}. An energetic initial state
radiation jet can be effectively used as a trigger for the event.  The
neutral states will obviously be missed completely. The charged particle
decay length and $\beta\gamma$ are also favourable for detection at the
MoEDAL detector \cite{moedal} at LHC. If observed, the masses of the
particles can be calculated from information about average energy loss
and reconstructed transverse momentum as measured from the curvature of
the charged tracks in the magnetic fields \cite{hqsp1,hqsp2}.

\begin{figure}[!h]
    \centering
    \includegraphics[width=0.45\textwidth]{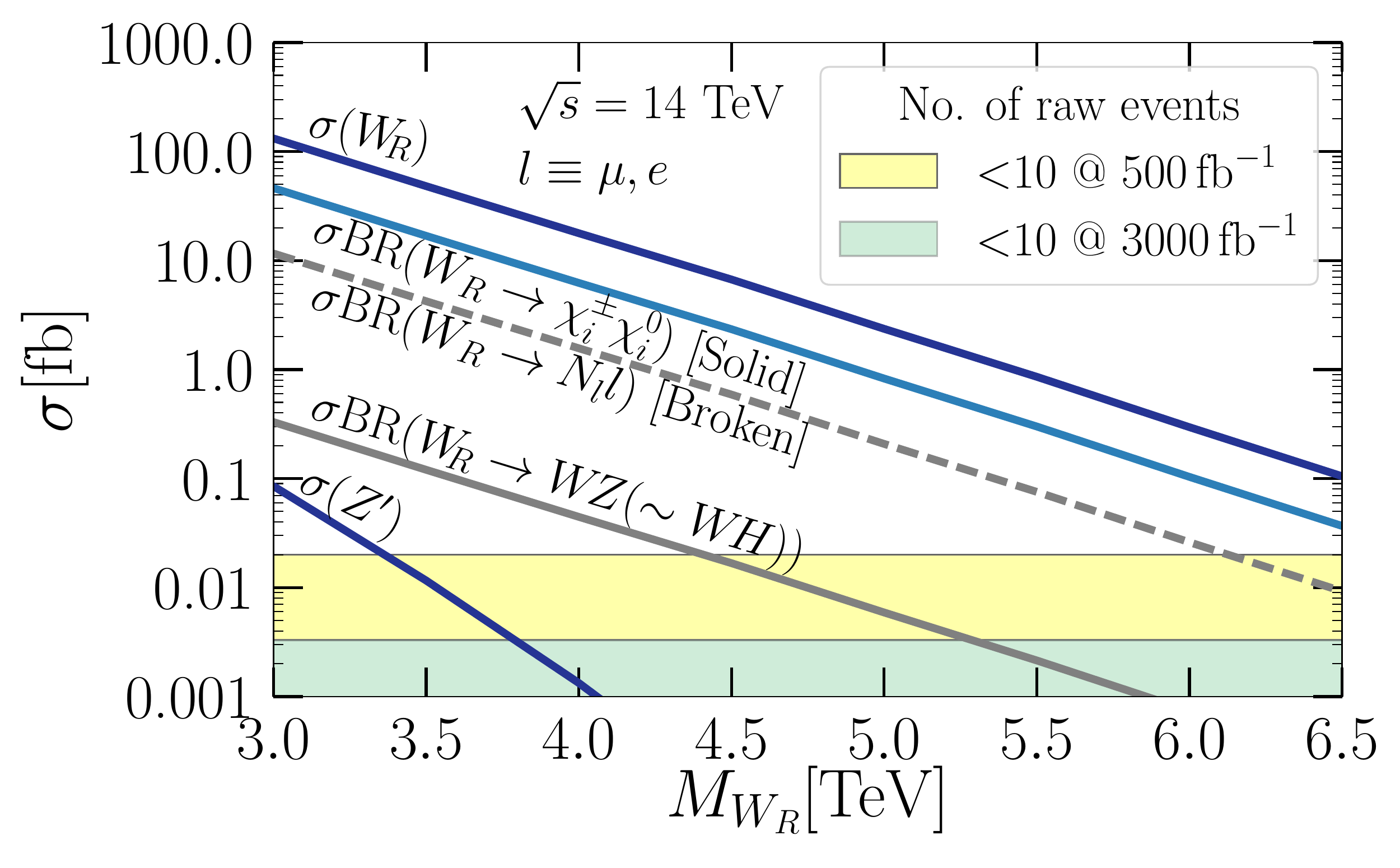}
	\caption{Cross section times branching fraction for production of
    $W_R$ and $Z'$ and their subsequent decays into different channels,
    as labelled, at $\sqrt{s}=14$  TeV. The thick lines represent total
    cross sections. The deeply (lightly) shaded regions
    delimit the cross sections for which the total
    number of raw events drops below 10 at 3000 (500) fb$^{-1}$.}
	\label{fig:cross}
\end{figure}

The vindication of TeV scale $SU(2)_R$ breaking will be the
discovery of the $W_R$ and the heavy neutrino, in the $lljj$
channel, the event topology being given by: $pp\rightarrow
W_R\rightarrow N_ll \rightarrow lljj$ \cite{ksp}. If the neutrino
is a Majorana particle as predicted by the LRS  model, one
should observe equal same-sign and opposite-sign final states.
We ask to what extent this signal is affected by the presence of
the self-conjugate triplets $\chi_{1,2}$?  In Fig.
[\ref{fig:cross}] we show the cross section times branching
fractions of $W_R$ production and its subsequent decays in
different channels\footnote{ The possibility of detecting a
virtual heavy $W_R$ signal through much lighter RH 
`neutrino jets' has recently been examined in \cite{manmit}}. For this purpose, leading order cross sections
were calculated in CalcHep 3.4 \cite{calchep} using the CTEQ6L1 parton
distribution functions \cite{cteq}, and multiplied by the
corresponding K-factors, as obtained from \cite{kfact}. For the
sake of comparison, we have chosen the DM multiplets to be mass
degenerate and having the smallest mass as allowed by relic
density constraints and taken $M_{N_l} = M_{\chi^\pm_{1,2}}$.

The dominant decay mode of $W_R$ is obviously to two jets. As can be
seen from Fig.  [\ref{fig:cross}] the decay $W_R^\pm \rightarrow
\chi_i^\pm \chi_i^0$ is a few times larger than the subdominant  but
often searched for leptonic decays ($l \equiv e, \mu$).  Nonetheless,
the leptonic branching remains substantial and we find that a $W_R$ with
mass $\sim 6.5$ TeV can still be discovered by the ATLAS and CMS
collaborations in this channel with $\sqrt{s} = 14$ TeV and an
integrated luminosity of 3000 fb$^{-1}$.  Indirect detection of a
heavier $W_R$ with masses up to $\sim 8$ TeV is possible in the studies
of K and B meson decays at LHCb \cite{bdb1} where this model has no
distinction from the canonical LRS model. Another promising mode for the
detection of $W_R$ is the di-boson channel  ($W_R \rightarrow W Z$ or
$W_R \rightarrow W H$). The branching ratios are almost the same for
these two channels. However, due to the suppressed $W_L$--$W_R$ mixing,
they are small, see Fig. [\ref{fig:cross}], and as $M_{W_R}$ approaches
6 TeV this channel becomes unfeasible.  Note that as the masses of the
triplet fermions are related to the mass of the $W_R$ boson from relic
density constraints, and since the $\chi_{1,2}^\pm$ do not interact with
the SM particles, the detection of $W_R$ or $Z'$ without detection of
these will essentially falsify the model.

With $M_{Z_R} \sim 1.94\times M_{W_R}$, the
discovery potential of $Z'$ is bleak at the LHC. For $W_R$ masses above
3.5 TeV, the $Z'$ becomes too heavy to be detected at LHC-II as can be
seen from Fig. [\ref{fig:cross}]. For HL-LHC luminosities of 3000 fb$^{-1}$,
the sensitivity increases slightly. 

{\em Conclusion:}
In this work we have presented a model which rests on left-right
symmetry, is amenable to gauge coupling unification, and provides
suitable dark matter candidates. Aided by two distinct discrete
symmetries inherent to the left-right symmetric theory the
stability of dark matter and the scales of symmetry breaking are
ensured. The model is falsifiable at both the GUT scale and the
LRS breaking scale at the Hyper-Kamiokande experiment and the LHC
respectively. The model predicts a `desert' between the LRS and
GUT scales.  In the absence of multiple symmetry breaking thresholds,
the variable parameters of the model viz. the $SU(2)_R$ coupling,
$g_R$, and scale of the $W_R$ mass are essentially fixed from
unification. The Dark Matter candidates satisfy the relic density
constraint aided by resonant enhancements of the cross section and
hence allowed masses are intertwined with $M_{W_R}/2$ and
$M_{Z'}/2$.  Their direct detection in ongoing and planned
experiments is unlikely. Nonetheless, with a very small leeway for the parameters to vary,
the model is remarkably predictive, making falsification or
vindication more or less straightforward at colliders.

{\bf Acknowledgements:} {\small TB acknowledges a Senior Research
Fellowship from UGC, India.  AR is partially funded by  the
Science and Engineering Research Board Grant No. SR/S2/JCB-14/2009.}

\printbibliography
\end{document}